\documentclass{aa}  
\usepackage{graphicx}
\usepackage{txfonts}
\begin{document}
   \title{XMM-Newton observation of the deep minimum state of PG~2112+059}
  \subtitle{A spectrum dominated by reflection from the accretion disk?}

   \author{N. Schartel 
          \inst{1} 
          \and 
          P.M. Rodr\'{i}guez-Pascual 
          \inst{1}
          \and 
           M. Santos-Lle\'{o} 
          \inst{1}  
          \and 
          L. Ballo 
          \inst{1}
          \and 
          J. Clavel 
          \inst{2}
          \and  
          M. Guainazzi 
          \inst{1}
          \and 
          E.~Jim\'{e}nez-Bail\'{o}n 
          \inst{3}
          \inst{4}
          \and 
          E.~Piconcelli 
          \inst{5}
       }          

   \offprints{N. Schartel}

   \institute{XMM-Newton Science Operations Centre, ESA, RSSD, ESAC, Apartado 50727,  
              E-28020 Madrid, Spain \\
              \email{Norbert.Schartel@sciops.esa.int} \\ \and
              Astrophysics Mission Division, ESA, SCT-SA, ESTEC, Postbus 299,
              2200 AG - Noordwijk, The Netherlands \\ \and
              Instituto de Astronom\'\i{}a, Universidad Nacional Aut\'onoma de M\'exico, 
              Apartado Postal 70-264, 04510-Mexico DF, M\'exico  \\ \and
               LAEFF-INTA, Apdo 50727, 28080-Madrid, Spain \\ \and
              Osservatorio Astronomico di Roma (INAF), via Frascati 33, 
                00040 Monteporzio Catone, Italy
             }

   \date{Received May 09, 2006; accepted August 06, 2007}

 
  \abstract
{Highly ionised absorbers and the frequent occurrence of relativistically broad iron 
fluorescence lines characterize the 0.2-10~keV spectra of (soft) X-ray weak quasars.  } 
{We constrain the physical conditions of the absorber and the broad iron 
line of the X-ray weak quasar \object{PG~2112+059} in greater detail than in previous studies.}
{We analyse a 75ks XMM-Newton observation of \object{PG~2112+059} performed in November 
2005 and compare it with a 15ks XMM-Newton observation taken in May 2003.}
{
\object{PG~2112+059} was found in a deep minimum state as its 
 0.2-12 keV flux decreased by a factor of 10 in comparison to the 
 May 2003 observation.  
During the deep minimum state the spectra show strong emission in 
 excess of the continuum in the 3-6 keV region. 
The excess emission corresponds to an $\rm{EW} = 26.1 \rm{keV}$
 whereas  its shape resembles that of heavily absorbed objects. 

The spectra of both observations of \object{PG~2112+059} can be 
 explained statistically by a combination of two absorbers where one shows a high 
 column density, $\rm N_{\rm H} \sim 4.5 \times 10^{23} cm^{-2}$, and the other 
 high ionisation parameters.
As the ionisation parameter of the high flux state, 
 $\rm \xi \sim 34  \;erg  \;cm \;s^{-1}$, is lower than the value found 
 for the deep minimum state, $\rm \xi \sim 110 \; erg \;cm\;s^{-1}$ , 
 either the absorbers are physically different or the absorbing material 
 is moving with respect to the X-ray source.

The spectra can also be explained by a continuum plus X-ray ionised 
 reflection on the accretion disk, seen behind a warm absorber. 
The ionisation parameter of the high state 
 ($\rm \xi \sim 5.6 \; erg \;cm\;s^{-1}$) is higher than the 
 ionisation parameter of the deep minimum state 
 ($\rm \xi \sim 0.2 \; erg \;cm\;s^{-1}$), as expected for a 
 stationary absorber.  
The values found for the ionisation parameters are in the range typical for AGNs.  
The spectra observed during the deep minimum state are reflection dominated 
 and show no continuum emission. 
These can be understood in the context of light bending near the supermassive 
 black hole as predicted by Minutti and Fabian.}
{
Light bending offers an alternative explanation for X-ray weak quasars and 
 might challenge the suggestion that absorption is the primary cause of their 
 X-ray weakness. 
If on a class level the weakness of X-ray weak quasars is caused by light 
 bending then they offer unique possibilities  to observe accretion disks near the 
supermassive black hole and even to test general relativity 
 in the strong field.}

   \keywords{quasars -- quasars indiviudal: \object{PG~2112+059}
                warm absorbers 
               }

   \maketitle

%

\section{Introduction}
About 10\% of quasars show an X-ray emission lower, by a factor  of 10-30, 
than expected based on their luminosity in other energy bands 
(Laor et al. \cite{Laor1997}; Wang et al. \cite{Wang1996};
Elvis \& Fabbiano \cite{Elvis1984}).  
These quasars are called X-ray weak or soft X-ray weak quasars.

Brandt, Laor and Wills  (\cite{Brandt2000}) showed that the optical to X-ray 
spectral index, $\alpha_{\rm OX} = 0.372 \log\left( f_{\rm 2 keV} / f_{3000 \AA} \right)$ 
 where $ f_{\rm 2 keV}$ and $f_{3000 \AA}$ are flux densities, 
 well separates soft X-ray weak from  ``normal'' 
 quasars, where soft X-ray weak quasars fulfil  $\alpha_{OX} \le -2$.  
X-ray weak quasars show a strong correlation between  $\alpha_{OX}$ and the 
C IV absorption equivalent width, which suggests that absorption is the 
primary cause of their soft X-ray weakness. 
The observed correlation between soft X-ray weakness and C IV absorption 
equivalent width is in general agreement with models, which invoke a connection 
between orientation and absorption strength. 
With increasing inclination angle we may observe quasars, non-BAL X-ray weak 
quasars, BAL quasars and type 2 quasars. 
Although the explanation of X-ray weakness as being due to absorption is 
 plausible, Brandt, Laor and Wills  (\cite{Brandt2000}) noted that a uniform 
screen of absorbing material covering both the X-ray source and the ultraviolet 
emission cannot reproduce the shape of the found correlation.
Either the UV and X-ray absorbers are distinct 
or the 
X-ray weakness is due to a peculiar property of the nuclear emission, such as 
extreme variability or unusual steepness.

Brandt, Laor and Wills (\cite{Brandt2000}) found several notable differences between 
X-ray weak and ``normal'' quasars in the Boroson \& Green (\cite{Boroson1992}) 
sample. 
Soft X-ray weak quasars are systematically characterised by low $[$O III$]$ 
luminosities and equivalent widths as well as distinctive H$\beta$ profiles. 
Most of them are located toward the weak  $[$O III$]$ end of the 
Boroson \& Green eigenvector 1, as are many BAL quasars. 
It has been suggested that unabsorbed Seyfert galaxies and quasars with 
eigenvectors 1s similar to that found for X-ray weak quasars are characterized 
by extreme physical parameters in the nuclear region, i.e. a high mass accretion 
rate relative to the Eddington limit ($\dot{M}/\dot{M_{Edd}}$).

The X-ray weak broad absorption line quasar \object{PG~2112+059} shows a complex 
 spectrum as well as a remarkable variability in the 0.2-10 keV region
 (Gallagher et al. \cite{Gallagher2004}; Schartel et al. \cite{Schartel2005}).
We use a deep XMM-Newton observation with a
 combination of spectral information and variability to constrain the physical
 properties of this quasar.
In this paper we present the analysis of the new XMM-Newton data from 2005.  
We compare the results with previous X-ray observations of \object{PG~2112+059} and 
 the XMM-Newton observation in 2002.
The paper is organized as follows: 
in Section~\ref{Sum} we provide a summary of the previous X-ray observations of \object{PG~2112+059}.
Both XMM-Newton observations and the performed analysis are described in 
 Section~\ref{OaDR}. 
The observed variability is presented in Section~\ref{Var}  
 and the spectral analysis in Section.~\ref{SPE}.  
We discuss our results in Section~\ref{Dis} and draw some 
 conclusions in Section~\ref{Con}.

\begin{table}
\caption[]{\label{TOBS} 
Exposure details of the November 2005 observation of \object{PG~2112+059}} 
\begin{tabular}{rcccc}
\hline
\\ 
N.$^{(1)}$ & In.$^{(2)}$ & Filter & Start         & Duration \\
          &        &        & Day\&Time     &             \\    
                 &        &        &  [UT]   & [ks] \\ \\  
\hline \hline \\
1  & M1 & thin 1          &  20 at 21:13:12 &  75.9  \\
2  & M2 & thick           &  20 at 21:13:12 &  75.9  \\
3  & pn & thin 1          &  20 at 21:35:54 &  74.2  \\	
6  & OM & V               &  20 at 21:17:50 &  3$\times$1.9 \\
9  & OM & U-NoBar         &  20 at 23:08:11 &  3$\times$1.9 \\
12 & OM & B               &  21 at 00:58:32 &  3$\times$1.9 \\
15 & OM & WHITE  	  &  21 at 02:48:53 &  3$\times$1.9 \\
18 & OM & VISIBLE GRISM 2 &  21 at 04:39:14 &  3$\times$4.0 \\	
21 & OM & UVW1            &  21 at 08:44:35 &  3$\times$1.9 \\
24 & OM & UVM2            &  21 at 11:04:56 &  3$\times$1.9 \\
27 & OM & UVW2            &  21 at 12:55:17 &  3$\times$1.9 \\
30 & OM & UV GRISM 1  	  &  21 at 14:45:38 &  3$\times$4.0 \\  	
\\ 
\hline \\
\end{tabular}
(1): exposure identifier; OM performed three exposures with each
          filter. Only for the first exposure is the exposure identifier
          provided.  
(2): instrument, where M1 stands for MOS 1 and M2 for MOS2.
\end{table}

\section{PG 2112$\rm+$059}
\label{Sum}
\object{PG~2112+059} is a member of the Palomar bright quasar survey 
 (Schmidt \& Green \cite{Schmidt1983}) and is located at a redshift of $z=0.456$ 
 (V\'eron-Cetty \& V\'eron \cite{Veron2000}).

The Hubble Space Telescope spectrum of \object{PG~2112+059}  (Jannuzi et al. \cite{Jannuzi1998}) 
 shows broad C IV absorption lines (Gallagher et al \cite{Gallagher2001}) 
 which classifies the source as a Broad Absorption Line (BAL) quasar. 
The quasar was observed with ROSAT in 1991 and based on the data, 
 Wang (\cite{Wang1996}) determined a low optical to X-ray spectral index 
 characteristic of X-ray weak quasars.

ASCA observed \object{PG~2112+059} in October 1999 (Gallagher et al. 
 \cite{Gallagher2001}) and in September 2002 the quasar was the target 
 of a Chandra observation (Gallagher et al. \cite{Gallagher2004}). 
The observations revealed a dramatic variability, where the ASCA observation 
 showed the highest flux, which was approximately a factor of four higher than 
 both in the earlier ROSAT observation and in the later Chandra observation.
An intermediate continuum flux state was measured with XMM-Newton in May 2003 
 (Schartel et al. \cite{Schartel2005}).

The spectrum of \object{PG~2112+059} taken in October 1999 with ASCA showed clear 
 indication of absorption but the statistics were insufficient to discriminate 
 between a neutral absorber, partial covering by a neutral absorber or an 
 ionised absorber. 
Neutral absorption could be excluded for \object{PG~2112+059} for the Chandra 
 spectrum taken in September 2002. An ionised absorber as well as a partially 
 covering neutral absorber are able to describe the Chandra data statistically 
 (Gallagher et al. \cite{Gallagher2004}). 
Both absorption scenarios require an increase of the absorbing column 
 density between 1999 and 2002.
On the other hand the XMM-Newton spectra of \object{PG~2112+059} from May 2003 
 require an ionised absorber. 
A neutral absorber as well as a partial covering absorption scenario can be 
 excluded from a statistical point of view  (Schartel et al. \cite{Schartel2005}).

\begin{table*}
\caption[]{\label{TSCR} 
Screening for low background level, source position and background extraction area }
\begin{tabular}{lcccccccccrr}
\hline \\ 
\multicolumn{1}{l}{N.$^{(1)}$} & \multicolumn{1}{l}{C.$^{(2)}$ } & \multicolumn{2}{l}{Source Position} 
    & \multicolumn{3}{l}{Screening Parameters$^{(3)}$ } 
    & \multicolumn{3}{l}{Background Extraction Area} 
    & \multicolumn{2}{l}{Net Source} \\
 &  & RA$^{(4)}$ &  Dec$^{(4)}$  &  Energy$^{(5)}$  & r$^{(6)}$ &  CR$^{(7)}$  &   
RA$^{(4)}$ &  Dec$^{(4)}$      & r$^{(8)}$ & CR$^{(9)}$  & $T^{(10)}$  \\  
 &  &     [h:m:s] &  [d:m:s]    & [$\rm keV$] & [$'$]  &  [$\rm s^{-1}$]  & [h:m:s] &  [d:m:s] & [$''$] & [$\rm 10^{-2} s^{-1}$] &
[$\rm ks$] \\  \\
\hline \hline \\ 
\multicolumn{12}{c}{Observation from 14.5.2003 / observation identifier 0150610201} \\ \\
1 & pn & 21:14:52.47 & 6:07:42.0 & 0.2-12.0 & 1-11 & 10. & 21:14:48.393 & 6:07:20.75 & 30 & 11.95$\pm$0.42 &   7.1 \\ 
2 & M1 & 21:14:52.44 & 6:07:41.5 & 0.2-10.0 & 1-14 & 5.0 &  source      & source & 40-105 &  3.50$\pm$0.18 &  10.8  \\
3 & M2 & 21:14:52.44 & 6:07:41.5 & 0.2-10.0 & 1-14 & 5.0 &  source      & source & 40-105 &  3.59$\pm$0.18 &  10.8  \\ \\
\multicolumn{12}{c}{Observation from 20.10.2005 / observation identifier 0300310201} \\ \\
1 & M1 & 21:14:52.60 & 6:07:41.0 & 0.2-10.0 & 1-14 & 2.0 & source       & source & 40-105 &  0.37$\pm$0.02 &  72.3 \\
2 & M2 & 21:14:52.60 & 6:07:41.0 & 0.2-10.0 & 1-14 & 2.0 & source       & source & 40-105 &  0.33$\pm$0.02 &  73.6 \\
3 & pn & 21:14:52.55 & 6:07:42.0 & 0.2-12.0 & 1-11 & 4.0 & 21:14:48.769 & 6:07:21.99 & 30 &  1.19$\pm$0.05 &  62.5 \\ 
\\  \hline \\
\end{tabular}

(1): exposure identifier;
(2): EPIC camera where M1 stands for MOS1 and M2 for MOS2;
(3): for the screening the data were binned with 100s; 
(4): in J2000; 
(5): energy range considered for the screening;
(6): the inner and the outer radius of the screening annulus;
(7): count rate threshold for rejection of screening;
(8): radius of background region (for annular background areas the inner 
        and the outer radius are provided);
(9): background corrected source count rate in the energy range of 
        0.2-12.0 keV for pn and 0.2-10.0 keV for MOS, respectively;
(10): accumulated exposure time.

\end{table*}

The October 1999 ASCA spectra of \object{PG~2112+059} showed no evidence for iron 
 line emission and Gallagher et al. (\cite{Gallagher2001}) determined an 
 upper limit for the narrow neutral iron line equivalent width of less 
 than 210 eV.  
Surprisingly, the spectrum of the quasar taken with Chandra in September 2002 shows a 
 broad iron line (Gallagher et al. \cite{Gallagher2004}). 
Such a broad line with the same width and the same 
 flux is below the detection limit of the ASCA spectrum from  1999 due to
 the much higher continuum flux.
Although the XMM-Newton spectra from May 2003 do not require a broad 
 iron line for a satisfying statistical description, adding a line with 
 the same width and flux as detected in the spectrum taken with 
 Chandra in September 2002 improves the description of the data 
 at the 97\% probability level (Schartel et al. \cite{Schartel2005}).  
The XMM-Newton spectra can be understood provided that the continuum is 
 increased with 
 respect to the Chandra observation; the broad iron line emission 
 is constant.

\section{Observations and data reduction}
\label{OaDR}

In 2005 XMM-Newton  (Jansen et al. \cite{Jansen2001}) observed \object{PG~2112+059} for 
about 76 ks. 
The observation started at 21 hours 12 minutes (UT) on the 20th of November 
 2005 and is archived under the observation identifier 0300310201 in the XMM-Newton 
 scientific archive.
In order to take advantage of the latest calibration progress and to allow an 
 optimal comparison between the two XMM-Newton observations we have re-analysed 
 the XMM-Newton data taken in May 2003.
Details of this observation are provided in Schartel et al. 
 (\cite{Schartel2005}). 
 This observation is archived under the observation identifier 0150610201.

During the XMM-Newton pointings three scientific instruments simultaneously 
observe: 
 the Reflection Grating Spectrometer (RGS; Brinkman et al. \cite{Brinkman2001}), 
 the European Photon Imaging Camera (EPIC) and the Optical Monitor (OM; Mason et al. \cite{Mason2001}). 
EPIC consists of three CCD cameras: the pn-camera  (Str\"uder et al. \cite{Strueder2001}) 
and two MOS-cameras (Turner et al. \cite{Turner2001}).  
All EPIC exposures of \object{PG~2112+059} were taken in the Full Frame mode with different 
 optical blocking filters in the light path. 
The details are provided in Table~\ref{TOBS}.
OM performed in total 28 observations with 
 different filters or grisms in the optical light path.  
All OM exposures were performed in the ``Science User Defined'' image mode.
The details are provided in Table~\ref{TOBS}.

We processed the data with the XMM-Newton Science Analysis System  
 (SAS) v. 7.0 (linux; compare Loiseau et al. \cite{Loiseau2006}) 
 with the calibrations from November 2006.

The pipeline products provided by the Survey Science Centre 
 (Watson et al \cite{Watson2001}) indicated that the RGS spectra of 
 \object{PG~2112+059} are either extremely weak or that the source is not detected at 
 all with RGS. 
We inspected the processed RGS events in the dispersion-cross-dispersion-plane 
 without finding any trace of a source spectrum for either of the two 
 XMM-Newton observations. 

We processed the pn and MOS data with the EPPROC and EMCHAIN routines of SAS, respectively.  
For pn single and double events were considered, which are characterized 
 by pattern 0 to 4. 
For MOS observations, events with pattern 0-12 were chosen, which correspond 
 to single, double, triple and quadruple events. 
The applied pattern selection follows the recommendation 
 (Kirsch et al. \cite{Kirsch2006}), ensures an 
 optimal energy calibration and allows standard detector response and 
 effective area calculation.

We followed the procedure of Piconcelli et al. (\cite{Piconcelli2005}) for 
 screening for time ranges with low radiation background level. 
The source extraction areas, the energy ranges used for the screening, 
 the corresponding rejection thresholds, the screening background extraction 
 areas and the final accumulated exposure times for each EPIC exposure are 
 given in Table~\ref{TSCR}.

The source events were extracted from a circular region with $ r=15^{''}$ 
 centred at the peak of the emission, which was determined by eye. 
Table \ref{TSCR} gives the source position used for the extraction.
The background was estimated from source-free regions near the source. 
For the pn-camera an offset circular region was chosen and for the MOS-cameras 
 an annulus around the source extraction region was selected.
The background for each exposure was determined from counts collected at the 
 same CCD in order to avoid the propagation of calibration differences. 
The centre of the circular background extraction region of the
  pn data was selected to have CCD y-coordinates similar to the source 
 extraction region, which minimizes effects resulting from the charge 
 transfer efficiency dependency on the detector coordinates.
All areas are specified in Table \ref{TSCR}.
The detector response and the effective area were generated with the 
 SAS routines RMFGEN and ARFGEN, respectively.

We processed the OM observations with the OMGCHAIN and OMICHAIN 
 routine of SAS, which applies all required astrometric and 
 photometric corrections. 
The two OM grisms were used in both observations of \object{PG~2112+059}.
In both observation the UV flux of \object{PG~2112+059} was too low 
 to accumulate useful spectra.
The spectra obtained with the visible grism in the light path 
 of OM are of sufficient quality.
The long observation of \object{PG~2112+059} allowed in addition OM 
 exposures with the U, V, B, UVW1, UVM2 and UVW2 filters in 
 the light path. 
For each filter three images were obtained.
We determined the fluxes through the mean of the three values 
 and provide the standard derivation as an estimate of their error. 
The obtained values are provided in Table \ref{tab:omflux}.

\section{Timing analysis}
\label{Var}

\subsection{OM}

The sequence of OM exposures in November 2005 allows us to set upper
limits to the variability of the source in different spectral bands on
timescales of a few hours. The average and the ratio of the
maximum to the minimum flux are given in Table~\ref{tab:omflux} for
each filter and show that the flux remains constant in all optical and
UV broad-band filters within a few percent.

Only OM Grism exposures were performed in the XMM observation on May
2003. Since \object{PG~2112+059} is rather faint in the UV for the OM UV-Grism,
we can only compare the optical spectra obtained with the V-Grism. The
optical spectrum obtained in November 2005 is systematically fainter
than the optical spectrum in May 2003 by $\sim$10\%. This difference
is only slightly larger than the measurement errors: $\sim$5\%.

\begin{table}[th]
\caption{Fluxes through OM filters}
\label{tab:omflux}
\begin{tabular}{lcc}
\\
\hline
\\
\multicolumn{1}{c}{Filter} &
\multicolumn{1}{c}{Flux} &
\multicolumn{1}{c}{Ratio} \\
\multicolumn{1}{c}{} &
\multicolumn{1}{c}{(1)} &
\multicolumn{1}{c}{(2)} \\
\\
\hline\hline
\\
V          &   2.08$\pm$0.01 &  1.01$\pm$0.02\\
B          &   2.91$\pm$0.04 &  1.03$\pm$0.01\\
U          &   3.83$\pm$0.01 &  1.00$\pm$0.01\\
UVW1       &   4.37$\pm$0.07 &  1.03$\pm$0.02\\
UVM2       &   4.76$\pm$0.06 &  1.03$\pm$0.04\\
UVW2       &   6.28$\pm$0.03 &  1.01$\pm$0.07\\
\\
\hline
\\
\end{tabular}

(1): Flux in units of 10$^{-15}$erg/cm$^{2}$/s/{\AA};
(2): Maximum to minimum flux ratio
\end{table}

\subsection{X-ray}

We based the X-ray variability analysis upon the pn observation 
 as this camera has the largest effective area of all X-ray 
 instruments onboard XMM-Newton. 
\object{PG~2112+059} shows a dramatic decrease of the X-ray flux between 
 May 2003 and November 2005: the 0.2 - 12 keV net source count rate dropped 
 by a factor of 10 from  0.116$\pm$0.004 s$^{-1}$ 
 to 0.0117$\pm$0.0004 s$^{-1}$.
The net source count rate is determined from the screened data. 
As the source shows in the November 2005 observation the lowest 
 flux of all X-ray observations reported so far, we refer in the 
 following to it as the ``deep minimum state'' of the source. 

The short time variability within the observation was tested based 
 on the events of the unscreened data. 
The source and background region were defined as described in Section~\ref{OaDR}. 
We generated source light curves with a binning time of 10.000s and 5.000s 
for the 350 eV - 500 eV, 500 eV - 2 keV, 2 keV  - 12 keV and 300 eV - 12 keV energy 
regions, respectively.  
We were not able to identify any time bin showing a significant derivation 
from the mean value. 
Based on this finding in combination with the size of the error bars, we 
 estimate that the source variability is below 20\% during the November 
 2005 observation.

\section{Spectral analysis}
\label{SPE}

We performed the spectral analysis of the MOS spectra over the energy 
range from 0.2 keV to 10.0 keV and of the pn spectra over the 0.2 to 
12.0 keV energy range, following the current 
recommendations (Kirsch \cite{Kirsch2006}).

In order to take full advantage of the progress in calibration and, 
 especially, to allow a rigorous comparison, we provide in the following 
 the spectral modelling for the deep minimum state from November 2005 as 
 well as of the high state from May 2003. 
In both cases we have added the two MOS spectra and have calculated the 
 corresponding effective areas and response matrices. 
We have done this for the November 2005 observation although the two 
 MOSs were exposed with different optical blocking filters in the light 
 path.
However, the accumulated counts at low energies are too few to take  
 advantage of the differences in effective area as originally intended  
 with the selection of the different optical blocking filters.

\begin{figure}[th]
 \centering
\includegraphics[width=6.5cm,angle=-90]{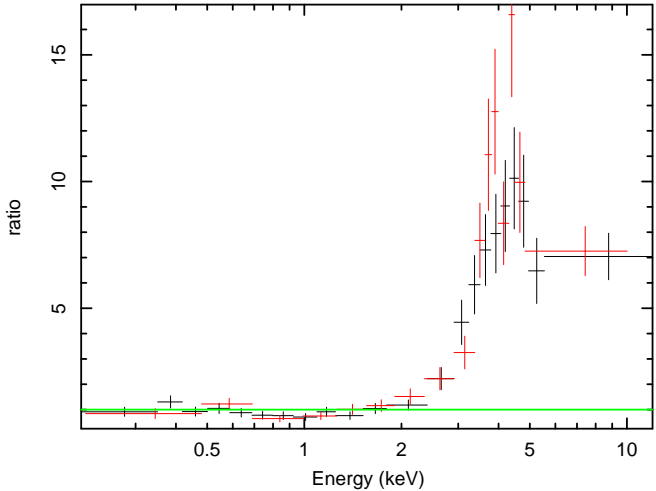}
\caption{The ratio between the EPIC data of the deep minimum state 
 of \object{PG~2112+059} and a hypothetical canonical power law spectra 
 with a photon index of $\Gamma =2$ behind Galactic absorption is shown.  
 The pn data points are given in black and the MOS data points are printed 
 in red, respectively.  }
 \label{Fig3}
   \end{figure}

The pn and (added) MOS spectra were binned with a signal-to-noise ratio 
of $>$5 restricted to a maximum of three bins per spectral resolution 
element. 
With this choice for the binning, over--sampling is excluded, a Gaussian 
distribution is approximately reached and $\chi^2$ statistics can 
be applied in the special fitting. 
The spectral analysis was performed with XSPEC 12.3.0c  
(Linux version, Arnaud \cite{Arnaud2006}). 
This software package uses the modified minimum $\chi^2$ method  (Kendall and 
Stuart \cite{Kendall1973}) for estimation of the spectral parameters, 
which requires applied binning.
The errors of estimated parameters are given for the 90\% confidence region 
for a single interesting parameter   ($\Delta \chi^2 = 2.7$, Avni \cite{Avni1976}).  

For all spectral models we assumed Galactic foreground absorption.
The Galactic equivalent column density in the direction of the quasar was 
 determined through pointed 21~cm radio observations to be
 $\rm N_H = 6.26 \times 10^{20} \rm cm^{-2}$ with an error of $~5$\%
 (Lockman \& Savage \cite{Lockman1995}).
For the different models the best estimated parameters, their errors and the 
 statistical parameter are provided in Table~\ref{TFITS};
 the same quantities for the emission or absorption components of the 
 models are reported in Table~\ref{TFITL}.

The spectral shape, observed during the deep minimum state of \object{PG~2112+059}, 
differs significantly from the shape observed during the high state in May 
2003. 
In the 3-6 keV region the minimum state shows a bump, like sources 
 that are partially covered by material with a high column density.
Figure~\ref{Fig3} shows the ratio between the EPIC data 
 of the deep minimum state and a hypothetical canonical power law spectrum 
 with a photon index of $\Gamma =2$ absorbed by material of Galactic column 
 density.   
This bump  
 was not present in the spectrum taken during the high state of the source. 
Given this appearing spectral feature we first tried to explain the spectra 
in terms of absorption and changes of it.

\begin{figure}[th]
 \centering
\includegraphics[width=6.5cm,angle=-90]{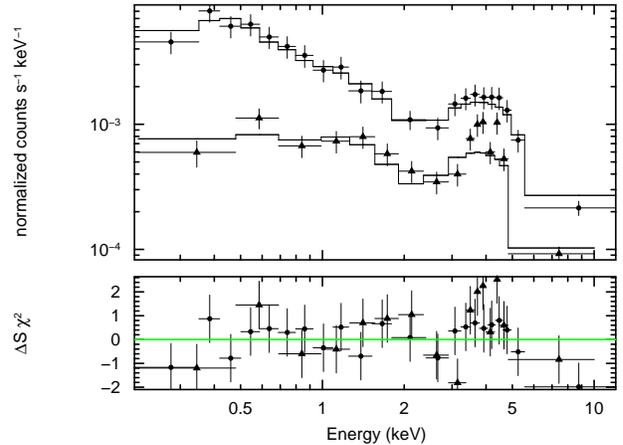}
\caption{The two (MOS and pn) EPIC spectra of the deep minimum state of \object{PG~2112+059} 
 are shown in comparison to a spectral model that consists of a continuum 
 power law behind two layers of absorbing material (Table~\ref{TFITS}, fit 10). 
 The upper spectrum was taken with the EPIC pn camera and the lower one with 
 the MOS cameras. 
The model is not able to describe the spectra well at energies above 3 keV as 
 there are systematic residua visible in the lower panel. }
 \label{Fig1}
   \end{figure}

\subsection{Continuum modeling with absorbers}
\label{cmwa}

Assuming a power law continuum we can exclude a neutral absorber 
  ($\rm \chi^2 = 108.3, d.o.f. = 51 $)
 as well as a neutral absorber that partially covers the source 
  ($\rm \chi^2 = 68.3, d.o.f. = 50 $)
 for the high state of 
 the source, whereas an ionised absorber allows a satisfying description of 
 the spectra (Table~\ref{TFITS}, fit 1).
For the deep minimum state a neutral absorber 
 ($\rm \chi^2 = 204.9, d.o.f. = 35 $) as well as ionised absorber ($\rm \chi^2 = 76.1, d.o.f. = 34 $)
 can be excluded, 
 whereas a neutral absorber partially covering the source provides 
a possible description of the source spectra (fit 8).  
As an ionised absorber cannot describe the deep minimum state of the source, 
 we can exclude that a change of the ionisation state of the absorbing 
 material in response to source variability explains the spectral differences.

\begin{table*}
\caption[]{\label{TFITS} Spectral fits of the EPIC spectra of \object{PG~2112+059}  }
\begin{tabular}{l@{ }l c@{ }c c@{ }c  c@{ }c c r@{ }r}
\hline \\ 
Fit & Model$^{(a)}$     & \multicolumn{4}{c}{Absorber}      
                        & \multicolumn{2}{c}{Continuum I} 
                        & \multicolumn{1}{c}{Continuum II} 
                        & \multicolumn{2}{c}{Statistic}  \\ 
       &                & \multicolumn{2}{c}{neutral}    
                        & \multicolumn{2}{c}{warm}   
                        & \multicolumn{2}{c}{power-law}
                        & \multicolumn{1}{c}{power-law}
                        & \multicolumn{2}{c}{ }  \\ 
    &                   & \multicolumn{1}{c}{$N_H^{(b)}$} & 
                          \multicolumn{1}{c}{$cf^{(c)}$}  
                        &  $N_H^{(b)}$ & $\xi^{(d)}$
                        & \multicolumn{1}{c}{$\Gamma$} & \multicolumn{1}{c}{$N^{(e)}$} 
                                                       & \multicolumn{1}{c}{$N^{(e)}$} 
                        & \multicolumn{1}{c}{$\chi^2$} & \multicolumn{1}{c}{d$^{(g)}$}   \\ 
    &                   & \multicolumn{1}{c}{{\small $[\rm 10^{22} cm^{-2}]$}} &
                           \multicolumn{1}{c}{ }  
                        & {\small $[\rm 10^{22} cm^{-2}] $} 
                        & \multicolumn{1}{c}{{\small $[\rm  erg \, cm \,s^{-1}] $}} & &  
                         \multicolumn{1}{c}{\small{$[\rm  keV^{-1}cm^{-2}s^{-1}]$}} & 
                         \multicolumn{1}{c}{\small{$[\rm  keV^{-1}cm^{-2}s^{-1}]$}} 
                        & \multicolumn{1}{c}{$ $} & \multicolumn{1}{c}{}   \\ \\ 
\hline \hline \\ 
\multicolumn{11}{c}{\underline{high state / observation from May 2003 }}\\ \\
1  &  $ ia * p $           &        &  
                              & $6.7_{-1.9}^{+2.3}$ & $87^{+24}_{-28}$
                              & $1.82_{-0.08}^{+0.14}$ & $1.2^{+0.2}_{-0.1}\; 10^{-4}$  & 
                                       &  44.7 & 50 \\  
2  &  $na * p + ia * p $  & $29_{-13}^{+18}$ &  $-$ 
                              & $3.2_{-1.2}^{+1.0}$ & $39^{+23}_{-16}$    
                              & $2.19_{-0.22}^{+0.12}$ & $1.4_{-0.9}^{+0.7} \; 10^{-4}$ 
                                & $1.2_{-0.2}^{+0.1} \; 10^{-4}$   
                                       &  35.4 & 48 \\   
3  &  $pa * ia * p $       & $28_{-13}^{+30}$ &  $0.51_{-0.20}^{+0.14}$   
                              & $3.3_{-1.2}^{+1.5}$ &  $39^{+22}_{-17}$   
                              & $2.14_{-0.16}^{+0.16}$ &  $2.4_{-0.7}^{+1.0} \; 10^{-4}$  & 
                                       &  35.1 & 48 \\   
4  &  $ia * ( p + ia * p) $     &                     &     
                              & $3.3_{-1.7}^{+1.5}$ &  $39^{-17}_{+22}$   
                              & $2.14_{-0.17}^{+0.16}$ &  $1.2_{-0.4}^{+0.2} \; 10^{-4}$  & 
                                       &     &   \\
   &                            &                     &     
                              & $35_{-20}^{+65}$ &  $2.4^{-2.4}_{+4900}$   
                              &                &  $1.2_{-1.0}^{+0.7} \; 10^{-4} $  & 
                                       & 34.9    & 47  \\  
5 & $(pa * ia * p) + ga$  & $26_{-13}^{+30}$ &  $0.49_{-0.23}^{+0.14}$  
                              & $3.3_{-1.2}^{+1.5}$ &  $39^{+22}_{-17}$  
                              & $2.14_{-0.17}^{+0.16}$ &  $2.3_{-0.7}^{+1.0} \; 10^{-4}$  & 
                                       &  34.2 & 47 \\   
6  &  $ ia * (p+bl(re)) $             &        &  
                              & $3.1_{-1.5}^{+1.6}$ & $35_{-22}^{+28}$  
                              & $1.90_{-0.10}^{+0.11}$ & $9.9_{-1.2}^{+1.8} \; 10^{-5}$  & 
                                       & 35.4 & 45 \\  
7 &  $ ia * (p+bl(re)) $             &        &  
                              & $2.9_{-1.4}^{+1.9}$ & $34_{-22}^{+32}$  
                              & $1.90_{-0.37}^{+0.18}$ & $9.7_{-2.3}^{+2.1} \; 10^{-5}$  & 
                                       & 36.0 & 46 \\   \\
\multicolumn{11}{c}{\underline{deep minimum state / observation from November 2005}} \\   \\
8   &  $ pa * p $            & $40_{-5}^{+6}$ & $0.96_{-0.17}^{+0.14}$
                              &       &   
                              & $2.23_{-0.07}^{+0.07}$ & $9.1_{-2.9}^{+0.7} \; 10^{-5}$ &  
                                       &  46.6 & 34 \\   
9   &  $na * p + ia * p $  & $46_{-7}^{+7}$ &  $-$ 
                              & $2.1_{-1.6}^{+0.9}$ & $74^{85}_{-53}$   
                              & $2.28_{-0.13}^{+0.07}$ & $1.0_{-0.4}^{+0.4} \; 10^{-4}$   
                               & $5.9_{-1.2}^{+1.1} \; 10^{-6}$ 
                                       &  40.4 & 32 \\   
10   &  $pa * ia * p $      & $44_{-7}^{+8}$  &  $0.94_{-0.03}^{+0.02}$  
                              & $2.2_{-1.6}^{+2.9}$  &  $80^{+105}_{-66}$   
                              & $2.25_{-0.19}^{+0.18}$  &  $9.8_{-3.1}^{+4.0} \; 10^{-5}$  & 
                                       &  40.1 & 32 \\ 
11  &  $ia * (p + ia *  p) $     &                     &     
                              & $2.3_{-1.8}^{+3.3}$ &  $87^{-70}_{+98}$   
                              & $2.20_{-0.19}^{+0.20}$ &  $5.7_{-1.3}^{+1.4} \; 10^{-6}$  & 
                                       &     &   \\
     &                            &                     &     
                              & $98_{(-32}^{+2}$ &  $91^{-66}_{+62}$   
                              &                &  $ 9.2_{-3.4}^{+5.1} \; 10^{-5}$  & 
                                       & 34.2    & 31  \\  
12   &  $(pa * ia * p) + ga $  & $42_{-7}^{+8}$  &  $0.93_{-0.04}^{+0.02}$   
                              & $2.4_{-1.8}^{+3.3}$  &  $88^{+98}_{-69}$   
                              & $2.21_{-0.19}^{+0.20}$  &  $8.5_{-1.9}^{+1.9} \; 10^{-5}$  & 
                                       &  33.5 & 31 \\ 
13$^{\dagger}$ & $ sw  * p $            & ($\sigma = 0.16^{+0.34}_{-0.04}$, &    $z=0.34_{-0.09}^{+0.32}$)
                              & $4.0_{-4.0}^{+3.6}$      &   $152^{+150}_{-152}$    
                              & $1.87^{+0.23}_{-0.11}$ & $1.3_{-0.2}^{+0.7} \; 10^{-4}$ &  
                                       & 42.6 & 48    \\    
14$^{\dagger}$ & $ pa * ed * p $            & $50_{-9}^{+14}$ & $0.97_{-0.07}^{+0.03}$   
                              &       &   
                              & $2.00^{f}$ & $9.0_{-1.6}^{+2.1} \; 10^{-5}$ &    
                                       &  16.0 & 14    \\   
15$^{\dagger}$ & $ pa * no * p $            & $55_{-11}^{+24}$ & $0.95_{-0.09}^{+0.05}$  
                              &       &   
                              & $2.00^{f}$ & $9.4_{-1.8}^{+3.1} \; 10^{-5}$ &   
                                       &  15.8 & 14    \\   
16$^{\dagger}$ & $ sw  * p $            & ($\sigma = 0.00^{+0.12}_{-0.00}$,        & $z=0.18_{-0.18}^{+0.04}$)
                              & $50_{-5}^{+0}$      &   $330^{+110}_{-100}$    
                              & $3.70^{+1.59}_{-0.64}$ & $1.3_{-1.3}^{+22} \; 10^{-3}$ &   
                                       &  21.1 & 13    \\    
17  &  $ na * (p+la) $             &   \multicolumn{2}{l}{$0.0_{-0.0}^{+2.8} \; 10^{-2} $} 
                              &  & 
                              & $1.97_{-0.21}^{+0.21}$  & $4.3_{-0.3}^{+0.4} \; 10^{-6}$  &   
                                       & 50.5 & 30   \\ 
18  &  $ ia * (p+bl(re)) $             &        &  
                              & $0.6_{-0.4}^{+0.3}$ & $0.01_{-0.01}^{+0.31}$ 
                              & $1.94_{-0.29}^{+0.22}$  & $0.0_{-0.0}^{+1.0} \; 10^{-6}$ &  
                                       & 29.9 & 29 \\  
19  &  $ ia * (p+bl(re)) $             &        &  
                              & $0.7_{-0.5}^{+0.3}$ & $0.03_{-0.03}^{+0.36}$ 
                              & $1.87_{-0.22}^{+0.29}$  & $0.0^{+2.1} \; 10^{-6}$ &  
                                       & 35.7 & 30 \\   \\
\multicolumn{11}{c}{\underline{both states$^{(g)}$ (high / deep minimum) }} \\   \\
20  &  $na * p + ia * p $  & $42_{-5}^{+9}$ &  $-$
                              & $3.01_{-0.9}^{+0.5}$ & $34^{+18}_{-12}$  
                              & $2.24_{-0.19}^{+0.07}$ & $2.0_{-1.3}^{+0.6} \; 10^{-4}$   
                                                       & $1.3_{-0.2}^{+0.1} \; 10^{-4}$
                                       &  & \\   
 &                             &             &   
                              &                & $110^{+82}_{-51}$   
                              & $2.20_{-0.16}^{+0.15}$  & $8.1_{-4.4}^{+3.0} \; 10^{-5}$   
                                                      & $6.0_{-0.9}^{+1.0} \; 10^{-6}$  
                                       & 78.8 & 82 \\   
21   &  $pa * ia * p $        & $43_{-6}^{+6}$ &  $0.58_{-0.10}^{+0.05}$  
                              & $3.1_{-0.8}^{+0.6}$ &  $34^{+14}_{-11}$ 
                              & $2.18_{-0.16}^{+0.07}$ &  $2.9_{-0.4}^{+0.4} \; 10^{-4}$  & 
                                       &   &  \\ 
    &                         &                         &  $0.94_{-0.03}^{+0.01}$  
                              &                         &  $106^{+90}_{-44}$ 
                              & $2.27_{-0.21}^{+0.05}$  &  $1.0_{-0.3}^{+0.1} \; 10^{-4}$  & 
                                       &  77.9 & 82 \\     
22  &  $ia * (p + ia * p) $     &                     &     
                              & $4.3_{-1.1}^{+1.3}$ &  $39^{-16}_{+23}$   
                              & $1.93_{-0.10}^{+0.09}$ &  $1.6_{-0.5}^{+0.4} \; 10^{-5}$  & &     &   \\
   &                            &                     &     
                              & $100_{-31}^{+0.0}$ &  $4430_{-1000}^{+570}$   
                              &                & $1.4_{-0.3}^{+0.8} \; 10^{-4}$    & 
                                       &    &   \\  
    &                        &                     &     
                              &                    &  $117^{-35}_{+216}$   
                              &                    &  $5.90_{-2.01}^{+1.07} \; 10^{-6}$  & &     &   \\
   &                            &                     &     
                              &                   &  $80.3_{-62.2}^{+27.5}$   
                              &                &   & 
                                       &  76.8  &  82 \\ 
23   &  $ia * (p+bl(re))$  &                         &    
                              & $0.9_{-0.2}^{+0.3}$     &  $5.7^{+3.1}_{-5.7}$ 
                              & $1.81_{-0.10}^{+0.11}$ &  $7.4_{-0.8}^{+0.9} \; 10^{-5}$  & 
                                       &   &  \\ 
    &                         &                         &    
                              &                         &  $0.2^{+0.7}_{-0.17}$   
                              &                         &  $0.0_{-0.0}^{+} \; 10^{-7}$  &
                              &  79.7 & 80 \\  

24   &  $ia * (p+bl(re))$  &                         &    
                              & $0.9_{-0.3}^{+0.3}$     &  $5.8^{+7.1}_{-3.3}$ 
                              & $1.80_{-0.09}^{+0.11}$ &  $7.3_{-0.8}^{+0.9} \; 10^{-5}$  & 
                                       &   &  \\ 
    &                         &                         &    
                              &                         &  $0.2^{+0.5}_{-0.17}$  
                              &                         &  $0.0_{-0.0}^{+8.7} \; 10^{-7}$  &
                              &  83.9 & 81 \\  
25   &  $ia * (p+bl(re))$  &                         &    
                              & $0.9_{-0.2}^{+0.3}$     &  $5.2^{+9.6}_{-3.3}$ 
                              & $1.79_{-0.16}^{+0.12}$ &  $7.2_{-1.0}^{+1.0} \; 10^{-5}$  & 
                                       &   &  \\ 
    &                         &                         &    
                              &                         &  $0.2^{+0.5}_{-0.2}$  
                              &                         &  $0.0_{-0.0}^{+7.9} \; 10^{-7}$  &
                              &  83.8 & 80 \\  
\hline \\
\end{tabular}
The errors are provided for the 90\% confidence level. 
Fits labeled with $\dagger$ were performed over the 2-10 keV energy range for MOS and the
2-12 keV energy range of pn.
Explanation of labels: (a) explanation of the used models:
$p$: power-law (XSPEC: pow), 
$na$: neutral absorber at the redshift of the source (XSPEC: zphabs), 
$pa$: partial covering with neutral absorber at the redshift of the source (XSPEC: zpcfabs), 
$ia$: ionized absorption at the redshift of the source  (XSPEC: absori),
$sw$: partially ionised absorbing material with large velocity shear (XSPEC: swind1)
$ed$: absorption edge at the redshift of the source (XSPEC: zedge),
$no$: notch line absorption  (XSPEC: notch),
$ga$: Gaussian line at the redshift of the source (XSPEC: zgauss),
$la$: relativistic disk line after Laor (XSPEC: laor, Laor \cite{Laor1997}),
$bl$: relativistic blurring from an accretion disk around a rotating black hole
      (XSPEC: kdblur, modified from laor model by Fabian and Johnstone),
$re$: reflection by a constant density illuminated atmosphere (XSPEC: Reflion, Ross \& Fabian \cite{Ross2005}).
Parameters: (b) equivalent column density, (c) covering fraction,
(d) ionization level of absorbing material $\xi=L/nr^{2}$,  as defined in {\em XSPEC},
 (e) photon flux at 1 keV, (f) fixed,
 (g) degrees of freedom.
 In all cases, where both states are fitted simultaneously, the best estimated 
 parameters are provided in two lines: 
 The first line provides the parameters for the high state and the second line for the 
 deep minimum state. In the second line, only parameters are given which were free to 
 vary with respect to the high state (first line). 
\end{table*}
\begin{table*}
\caption[]{\label{TFITL} Spectral fits of the emission/absorption features of PG~2112+059}
\begin{tabular}{llclcccccc}
\hline \\ 
Fit & Emission   &  Energy$^{(b)}$   & Parameter$^{(c)}$ & Norm$^{(d)}$ & R(in) & Index & Inclination$^{(e)}$  & 
$\Xi^{(g)}$ & \\ 
    & Line Model &  [$ \rm keV$]  &  & [$ \rm cm^{-2}\;s^{-1}$]  &    [$\rm G\;M\;c^{-2}$]   &      & [$\rm  degree$]  
&  [$\rm erg \, cm \,s^{-1}$] & \\ \\
\hline \hline \\
\multicolumn{10}{c}{\underline{high state / observation from May 2003 }}\\  \\
5    & ga            &  $6.40^{f}$ & $W = 0.0^{f} eV $ & $1.4_{-1.4}^{+2.5} \;10^{-6}$   
                    &       &  & &    \\  
6   &  bl(re)  & &    &  $6.0_{-5.2}^{+3.9} \;10^{-7}$  &    
                      $5.5^{+95}_{-4.3}$  & $ 10.0_{-6.4}^{+0.0}$  & $34^{-32}_{+22} $ &
                      $31_{-31}^{+84}$  & \\   
7   &  bl(re)  & &      &  $6.4_{-6.3}^{+8.3} \;10^{-7}$  &    
                      $1.235^{f}$  &  $ 2.9_{-2.9}^{+7.1}$  & $21^{-21}_{+69} $ &
                      $31_{-31}^{+59}$  & \\  
\\
\multicolumn{10}{c}{\underline{deep minimum state / observation from November 2005}} \\  \\
12    & ga            &  $6.40^{f}$ & $W = 0.0^{f} eV $ & $5.0_{-3.2}^{+3.2} \;10^{-7}$    
                    &       &  & &    \\  
14$^{\dagger}$   & ed         & $7.8_{-0.3}^{+0.5}$  & $ \tau = 1.3_{-0.5}^{+0.6} $ &     &  & & & &   \\  
15$^{\dagger}$   & no         & $8.2_{-0.6}^{+0.4}$  & $ cf = 0.45_{-0.15}^{+0.13} $ &     &  & & & &   \\  
17   & la          & $4.0_{-0.1}^{+0.1} $  &  ($EW = 26.1_{-2.8}^{+2.6} keV$) &  $7.1_{-0.8}^{+0.7} \;10^{-6}$  & 
                     $1.6_{-0.2}^{+0.3}$ & $ 2.8_{-0.4}^{+0.5}$ & $86.6_{-0.3}^{+1.8} $ &   \\  
18   &  bl(re)  & &          &  $5.7_{-2.5}^{+4.7} \;10^{-7}$  &  
                      $1.9^{+0.5}_{-0.5}$  & $ 6.2_{-3.9}^{+3.8}$ & $58_{-21}^{+11} $ &
                       $30_{-30}^{+11}$ & \\  
19   &  bl(re)  & &        &  $4.1_{-1.3}^{+5.4} \;10^{-7}$  &  
                      $1.235^{f}$  & $ 2.9_{-0.2}^{+0.3}$   & $39_{-7}^{+9} $ &
                       $34_{-34}^{+11}$ & \\  \\
\multicolumn{10}{c}{\underline{both states$^{(h)}$ (high / deep minimum) }} \\  \\
23   &  bl(re)  &       &    &  $6.3_{-2.2}^{+3.6} \;10^{-7}$  &   
                      $2.7^{-0.6}_{+0.9}$  & $3.2_{-0.4}^{+3.1}$   & $40_{-7}^{+14} $ 
                                & $37_{-4}^{+9}$  & \\ 
    &           &       &       &  $3.7_{-0.8}^{+1.2} \;10^{-7}$  &  
                        &       &  &  &  \\
24   &  bl(re)  &       &     &  $6.4_{-2.1}^{+3.6} \;10^{-7}$  &   
                      $1.235^{f}$  & $ 2.8_{-0.2}^{+0.2}$  & $38_{-7}^{+6} $ &
                       $36_{-4}^{+8}$  & \\ 
    &           &       &       &  $3.6_{-0.7}^{+1.2} \;10^{-7}$  &  
                        &       &  &  &  \\
25   &  bl(re)  &       &     &  $5.8_{-5.4}^{+5.0} \;10^{-7}$  &   
                      $1.235^{f}$  & $2.8_{-0.2}^{+0.2}$  & $38_{-7}^{+6} $ &
                       $40_{-40}^{+151}$ &  \\ 
     &  bl(re)  &       &                      &  $3.6_{-0.9}^{+1.2} \;10^{-7}$  &   
                                          &           &  &
                                                  $37_{-4}^{+8}$ & \\ 
\\
\hline \\
\end{tabular}

The corresponding continuum fits are provided in Table \ref{TFITS}.
Fits labeled with $\dagger$ were performed over the 2-10 keV energy range for MOS and the
2-12 keV energy range for pn.
(a) The following models are used:
$ed$: absorption edge at the redshift of the source (XSPEC: zedge),
$no$: notch line absorption  (XSPEC: notch),
$ga$: Gaussian line at the redshift of the source (XSPEC: zgauss),
$la$: relativistic disk line after Laor (XSPEC: laor, Laor \cite{Laor1997}), maximal outer radius R$_{out}$ = 400 ,
$bl$: relativistic blurring from an accretion disk around a rotating black hole
      (XSPEC: kdblur, modified from laor model by  Fabian and  Johnstone), 
      maximal outer radius R$_{out}$ = 400 and iron abundance to iron/solar=3,
$re$: reflection by a constant density illuminated atmosphere (XSPEC: Reflion, Ross \& Fabian \cite{Ross2005}). 
Parameters:
 (b): energy of line in rest frame of quasar, except for fit 17, 
 (c): open parameter,
 (d): photon flux, 
 (e): inclination angle,
 (f): fixed, 
 (g): ionisation parameter.
 In all cases, where both states are fitted simultaneously, the best estimated 
 parameters are provided in two lines: 
 The first line provides the parameters for the high state and the second line for the 
 deep minimum state. In the second line, only parameters are given which were free to 
 vary with respect to the high state (first line). 
\end{table*}

In order to test for more complex absorption scenarios, we tried first a model 
 assuming that part of the light is absorbed by neutral material 
 and that the remaining part  of the light is absorbed by ionised material. 
This double-absorber scenario provides a statistically significant description of the 
 spectra obtained during the high state (fit 2) as well as during the deep 
 minimum state of the source (fit 9).  
During the high state the covering fraction is 54\% for the neutral material 
 ($ \rm N_H = 29_{-13}^{+18} \times 10^{22} cm^{-2}$) and 46\% for the ionised material 
 ($ \rm N_H = 3.2_{-1.2}^{+1.0} \times 10^{22} cm^{-2}$). 
The corresponding numbers for the low state are 94\% 
 ($ \rm N_H = 46_{-7}^{+7} \times 10^{22} cm^-2$) and 6\% 
 ($ \rm N_H = 2.1_{-1.6}^{+0.9} \times 10^{22} cm^{-2}$).
The absorption scenario even allows a description of both states assuming 
 constant column densities but allowing different ionisation states and 
 relative normalization of the continuum components (fit 20), 
 where the latter equals  a change of the covering factors.
The obtained high state best-fit values correspond to a covering fraction of 61\% 
 for the neutral absorber 
 ($ \rm N_H = 42_{-5}^{+9} \times 10^{22} cm^{-2}$) and 39\% for the ionised material 
 ($ \rm N_H = 3.01_{-0.9}^{+0.5} \times 10^{22} cm^{-2}$).
The joint fit of both states estimates the flux of the high state to be 3.8 
times the flux emitted during the deep minimum state. 
Hence assuming constant location and density of the absorbing material we expect 
 the ionisation parameter of the high state to be  3.8 times higher
  than the ionisation parameter of the deep minimum state. 
This is in sharp contrast to the observations, which show instead a high state 
 ionisation parameter 3 times lower than the low state one.

Similarly we tested a scenario with an ionised absorber and a neutral absorber 
that partially covers the source in the light path to the X-ray source. 
This scenario implies a layer of material that is located sufficiently near 
to the continuum source to be ionised and clouds of neutral material.
This scenario provides an acceptable statistical description of the spectra 
taken during both states of the source (fit 3 and fit 10).
Again the scenario provides a statistically acceptable description of both states 
 assuming constant column densities.  
However, it requires a change of the covering factor of the neutral absorber 
 beside the change of the ionisation states (fit 21).
Similar to the previous scenario we can compare the variability with the 
 change of the ionisation parameter. 
Assuming a constant location and density of the absorbing material we would expect 
 that a flux variability of a factor of $\sim$3 implies the ionisation parameter of the 
 high state to be a factor of 3 higher than the ionisation parameter of the deep 
 minimum state. 
Again the opposite is the case: the ionisation parameter in the high state is a 
factor of 3 lower than the low state.

Next we considered the possibility that \object{PG~2112+059} is covered by an ionised 
 absorber partially covering the X-ray source in combination with a further 
layer of ionised material that covers the X-ray source completely. 
This scenario allows a satisfying description of the two observations individually 
(fit 4 and 11) as well as of the two observations together (fit 22). 
In the latter case a high column density absorber  covers about 90\% of the 
 X-ray source and a low density absorber covers the entire source. 
Again, it is interesting to check the behaviour of the ionization parameters of 
 the two absorbers as the source flux changes. 
As before for the low-column-density absorber, the ionisation parameter in the 
 high state is a factor  of 3 lower than in the deep minimum state whereas we would 
 have expected a drop by a factor of three. 
For the high-column-density absorber the change in the ionisation parameter goes 
 in the direction of the flux variability but significantly exceeds the expected 
 value: the observed ratio of the ionisation parameters is 55 in contrast with a 
2.6 flux ratio as expected.

In all three discussed complex absorption scenarios the ionization of the absorber 
 does not follow the flux variation as expected. 
This inconsistency is not an artefact of fitting 
 both states simultaneously (fit 20, fit 21 and fit 22). 
The inconsistency is present even when the two states are fitted separately with 
 different column densities (fit 2 and fit 9, fit 3 and fit 10, fit 4 and fit 11) 
 although errors are clearly larger, which prevents firm conclusions.

Broad absorption line quasars, like \object{PG~2112+059}, are characterized by 
 outflows with significant velocities. 
Based on HST observations Gallagher et al. (\cite{Gallagher2004}) determined the outflow velocities 
 of four variable UV absorbing features in the spectrum of \object{PG~2112+059} 
 to be 9,900, 13,500, 16,500 and 20,100 km/s. 
Therefore we considered the possibility that the X-ray absorbers are also present 
 in the form of powerful winds with velocity gradients as represented in the 
 ``SWIND'' model (Gierli{\'n}ski \& Done \cite{Gierlinski2006}, 
                 Gierli{\'n}ski \& Done \cite{Gierlinski2004}). 
The model allows a fair description of the high state of \object{PG~2112+059}  
 (fit 13) but is not able to provide a valid 
 description of the deep minimum state 
 ($\rm \chi^2 =86.3$, $\rm d.o.f. = 32$)
 and as such the scenario was not considered further
 for the entire energy band.

\subsection{Lines, bumps and reflection models}
\label{lbarm}

We chose the absorption scenario that is composed of a partial covering neutral 
absorber and a total covering ionized absorber  (i.e. fit 10) 
to determine upper limits for the presence of  a narrow neutral iron emission line. 
To do this we added a Gaussian line profile to the model with the position and the 
 width fixed to 6.4 keV and 0.0 eV, respectively, leaving the normalization as 
 a free parameter. 
For the high state of the source (fit 5) inclusion of a Gaussian line reduces 
 the value of $\chi^2$ by $\Delta \chi^2=0.9$
(3-$\sigma$ upper-limit to $\rm EW_{3\sigma} \le 559 \;\rm eV$).
For the deep minimum state (fit 12) the value of the $\chi^2$ decreases by 
 $\Delta \chi^2=6.6$, corresponding to a 2.6-$\sigma$ detection of the line 
 with an equivalent width of $\rm EW = 160^{+100}_{-110}\;\rm eV$.
The corresponding 3-$\sigma$ upper-limit is $\rm EW_{3\sigma} \le 354 \;\rm eV$.
As the fit without a Gaussian line shows systematic residua over the entire bump region 
 we cannot consider the improvement in $\chi^2$ as an indication of  the presence of 
 a narrow  neutral 6.4 keV iron line.

\begin{figure*}[th]
 \centering
\includegraphics[width=13cm,angle=-90]{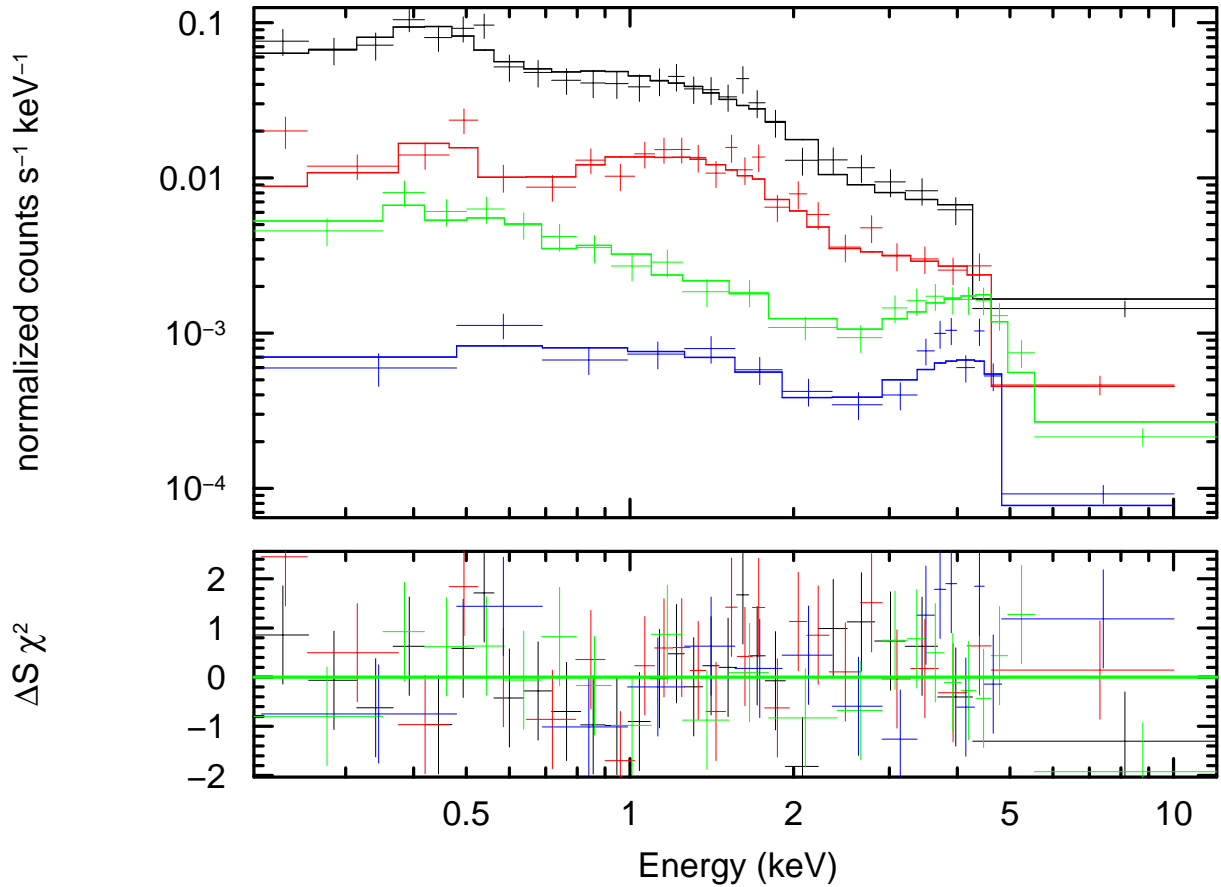}
\caption{
The EPIC spectra of both observations plotted in comparison to a model 
 consisting of a continuum power law and a relativistically blurred X-ray ionised 
 reflection component that originates in an accretion disk near to a supermassive 
 black hole. 
Both components are seen through a warm absorber (Table~\ref{TFITS}, fit 23). 
The ionisation parameter of the warm absorber and the normalization of the continuum 
 components were free to vary. 
The deep minimum state is dominated by the reflection component while the power 
 law component vanishes. }
 \label{Fig2}
   \end{figure*}

Although the discussed absorption scenarios provide statistically acceptable 
 descriptions of the spectra, the models fail to provide a good 
 description of the bump observed during the deep minimum state as there are 
 always residua left, compare Figure~\ref{Fig1}. 
These residua are too large and too systematic in character to be explained 
 with current calibration uncertainties.

In order to study the nature of the 3-6 keV bump of the deep minimum state in detail 
we restricted the spectral analysis to the 2-10 keV energy range of the spectra. 
As expected the spectra can be described with a power law continuum and a neutral 
absorber that covers the continuum source partially. 
The power-law index of the continuum is estimated to be $\Gamma=3.54_{-0.57}^{+0.43}$. 
As changes of the covering factor essentially only impact the low energy region, 
 the power-law index is mainly determined by the spectral shape at energies 
 $>$6 keV in the rest frame of the source.  
But a power-law index of this value is rather unusual for AGN in this energy region
 (Piconcelli et al. \cite{Piconcelli2005}). 
We therefore modelled the spectra with a partially covering neutral absorber with 
 an additional absorption edge assuming a power law continuum with a photon index 
 fixed to the canonical $\Gamma=2.0$ (fit 14). 
The estimate for the energy of the absorption edge, $\rm E=7.78^{+0.78}_{-0.31} \rm keV$, is above the value 
 of neutral iron. 
However, at the 3-$\sigma$ level, absorption by neutral iron cannot be excluded
 ($\rm E_{3\sigma}=6.92 \rm keV$). 
In addition we tested a model in which the edge was replaced by an absorption 
 feature with the shape of a notch (fit 15). 
The lower border of the notch was estimated to be $\rm E_{\rm low}=7.59^{+0.51}_{-0.87} \rm keV$, 
 where the values are provided in the rest frame to the quasar. 
Again an absorption by neutral iron cannot be excluded ($\rm E_{\rm low / 3\sigma }=6.40 \rm keV$). 
To summarize, the analysis of the high energy range of the spectra shows that the residua can 
 be reduced by either assuming an uncommon steep continuum at energies 
$>$6 keV or an additional, third, absorption feature. 
If the absorber is identified with iron then ionised material would be preferred, 
 but neutral iron cannot be excluded.

As an absorbing edge as well as a notch are rather simple and likely unrealistic models we 
 tried two, more sophisticated absorption scenarios to test if the bump can be 
 interpreted as an artefact of an insufficient description of absorption. 
We use the deep minimum spectra limited to the 2-10 keV energy 
 range and explored models that allow one to account for winds with velocity dispersions.
First we tried the ``SWIND'' (Gierli{\'n}ski \& Done \cite{Gierlinski2006}, 
                 Gierli{\'n}ski \& Done \cite{Gierlinski2004})
 model.  
A ``SWIND''-absorbed power law continuum is able to describe the 2-10 keV spectra 
 from a statistical point of view (fit 16), although the obtained $\rm \chi^2$-value is 
 rather high, $\rm \chi^2 = 21.12$   with respect to $\rm d.o.f. = 13$.  
But most important, the best estimate of the photon index of the power law continuum,
 $\rm \Gamma=3.70^{+1.59}_{-0.64}$ is too high for the modelled energy range. 
An attempt to describe the spectra with a ``SWIND''-absorber and the photon index 
 of the continuum fixed to the canonical value, $\rm \Gamma_{canonical} = 2.0.$ 
 failed ($\rm \chi^2 = 39.7$ for $\rm d.o.f. = 14$). 
The second complex absorption model that we tried is XSTAR (Kallman \cite{Kallman2007}). 
Within the statistical analysis program, XSPEC, the XSTAR model is provided as 
 pre-calculated tables as well as in analytical form. 
Currently, the analytical from is not fully tested and requires a substantial amount of 
 computing power, but offers the advantage that more parameters are allowed to be varied 
 than in the tables. 
Our attempts to model the 0.2-10 keV spectra of the deep minimum state with the XSTAR 
 table models failed.    
Statistically acceptable fits could only be obtained at the expense of unrealistically 
 high photon indices, e.g. $\rm \Gamma=6.77^{+0.55}_{-0.63}$ for $z=0.457$ 
 ($\rm \chi^2 = 20.4$ for $\rm d.o.f. = 15$) or 
 $\rm \Gamma=4.82^{+0.35}_{-0.31}$  for $z=0.457_{-0.457}^{+0.0}$
 ($\rm \chi^2 = 20.5$ for $\rm d.o.f. = 14$).  
Attempts to model the spectra with a canonical photon index, $\rm \Gamma_{canonical} = 2.0.$ , 
 were statistically unacceptable ($\rm \chi^2 = 37.4$ for $\rm d.o.f. = 15$ for $z=0.457_{-0.457}^{+0.0}$
 and $\rm \chi^2 = 37.4$ for $\rm d.o.f. = 16$ for $z=0.457$).  
Trials with the analytical XSTAR model confirmed the results of the table models. 
Either the fits are statistically unacceptable or unrealistic parameters were estimated. 
The possibility to vary the iron abundance could not be utilized to improve the situation 
 as unrealistically high values are estimated. 
Again, we always cross-checked the results by trying fits with parameters fixed to the 
 canonical values, e.g. $\rm \Gamma_{canonical} = 2.0.$ and iron/solar=3,
 and the models were statistically unacceptable.

Since absorption does not seem to be the best possible interpretation of 
 the 3-6 keV bump observed during the deep minimum state, we tried to model 
 the spectra  (without limiting to the hard emission) with a power 
 law continuum absorbed by ionised material plus (1) an additional Gaussian line or 
 (2) a relativistic broadened emission line emitted by an accretion disk around a 
 Kerr black hole (Laor \cite{Laor1991}). 
The parameter estimates result in both cases in very high column densities 
 ($\rm N_H = 6.6_{-1.9}^{+2.9} \times 10^{23} \; \rm cm^{-2}$ and 
 $\rm N_H = 9.4_{-4.0}^{+0.6} \times 10^{23} \; \rm cm^{-2}$) for the 
 ionised absorber in combination with very high ionisation parameters 
 ($\xi = 670_{-140}^{+310} \;\rm erg \;cm\;s^{-1}$ and $\xi = 884_{-390}^{+750} \;\rm erg \;cm\;s^{-1}$) and 
 a very  flat (Piconcelli et al. \cite{Piconcelli2005}) power 
 law continuum ($\Gamma = 1.37_{-0.10}^{+0.11} $ and $\Gamma = 1.30_{-0.06}^{+0.08}$). 
Given these parameter estimates, in both cases the warm absorber is involved in at least 
 part of the 3-6 keV bump emission, whereas the continuum is unusually flat.

Next we tried to describe the continuum with a power law absorbed by neutral material. 
In the case of a description of the bump through a Gaussian line, the power law 
 continuum is estimated to be unusually steep, $\Gamma = 2.67_{-0.51}^{+0.37}$ in 
 comparison with other quasars (Piconcelli et al. \cite{Piconcelli2005}).  
The steep continuum leads to a severe underestimation of the continuum contribution for 
 the region of the bump which consequently causes an unrealistically wide equivalent width. 
Therefore the scenario was not considered further. 
The situation looks better if the bump is modelled with a relativistic broadened 
 emission line (fit 17), where we fixed the outer radius of the disk. 
The best estimated power law index, $\rm \Gamma = 1.95_{-0.20}^{+0.22}$, is  
 in the range expected for quasars but we formally obtain 
 $\rm EW = 26.1_{-2.8}^{+2.6}  \; \rm keV$.
The model to data comparison shows systematic residua and the fit does not appear 
 to be statistically very satisfying ($\rm \chi^2 = 50.7, d.o.f. = 30 $). 
Such a value of the  equivalent width indicates that the bump is by far dominated by 
 the line emission for energies above $\rm 3 keV$. 
From a physical viewpoint the obtained equivalent width shows that the neutral absorber plus 
 an emission line model cannot 
 account for the  measured spectrum  and that especially the reflection continuum 
 must be considered accordingly. 

Consequently we tested whether the deep minimum state is able  to be interpreted by X-ray 
 ionised reflection on the accretion disk. 
We assumed a continuum spectrum composed of a power law and an ionised-reflection 
 model  (Ross \& Fabian \cite{Ross2005})  that is blurred with a Laor line profile 
 (Ross \& Fabian \cite{Ross2005};  Crummy et al. \cite{Crummy2005}).  
We model the data with this continuum and an ionised absorber in the line of sight.  
We restricted the free parameters of the warm absorber by assuming a temperature 
 of T=3$\times$10$^5$K and solar iron abundance.
In addition we assumed that the power law component and the reflection model 
 have the same, but free, index. 
This model allows an excellent statistical description of the data although the 
 parameters of the relativistic blurring cannot be well constrained. 
The best parameter estimates either provide a very low value for the outer 
 radius of the emission region of the accretion disk in combination with very 
 high iron abundance relative to the solar abundance or a very steep index for 
 the power law emissivity.  
We therefore fixed in all models the outer radius of the disk emission region to 
 its maximal radius R$_{out}$ = 400 and the iron abundance to iron/solar=3.  
Modelling of the spectra with either the inner radius of the emission region 
 as a free parameter (fit 18) or fixing the inner radius to its minimal 
 possible value R$_{in}$ = 1.235 (fit 19) result in acceptable descriptions 
 ($\rm \chi^2= 29.9$ for $\rm d.o.f. = 29$  and $\rm \chi^2 = 35.7$ for 
 $\rm d.o.f. = 30$) of the data obtained during the deep minimum state of 
 the source. 
For completeness we verified that the described model is also able to describe 
 the high state of the source (fit 6 and fit 7). 

Finally we tested whether a model that consists of a power law and a relativistically 
 blurred ionised reflection behind an ionised absorber can explain both 
 observations of \object{PG~2112+059} in a consistent and simple way.  
To do this we fitted both data sets simultaneously. 
Again we reduced the number of free parameters by fixing parameters 
 as specified above. 
In addition we coupled the large majority of the parameters such that the fit 
 was forced to find an estimate describing both observations.  
Between the two observations only three parameters were allowed to vary: the 
 normalizations of the two continuum components and the ionisation parameter 
 of the warm absorber. 
The model provides an excellent description of the data; no 
 systematic residua are left over if the model is compared with the data 
 (Figure \ref{Fig2}). 
Similar to the modulations described in the previous paragraph, it was necessary 
 to constrain the relativistic blurring by fixing the iron abundance to three 
 times the solar value and the outer radius of the disk emission region 
 to R$_{out}$ = 400. 
Both fits, with free inner radius (fit 23) and with the inner radius fixed to 
 the minimal possible inner radius (fit 24) provide statistically excellent 
 descriptions of the data ($\rm \chi^2= 79.9$ for $\rm d.o.f. = 80$  and $\rm \chi^2 = 83.9$ for 
 $\rm d.o.f. = 81$)  without showing systematic residua.

The limited statistics required a freezing of two parameters that describe 
 the ionised-reflection model blurred with a Laor line profile and therefore 
the estimated parameters cannot be interpreted in a simple way. 
Nevertheless the models provide a plausible description of the data which is 
 in agreement with general expectations and measurements.

\section{Discussion}
\label{Dis}

A simple model, consisting of a power law continuum behind a layer 
 of absorbing material, fails to describe both states of 
 \object{PG~2112+059} in a consistent way.
The upper limits obtained for a possible narrow iron fluorescence 
 emission line are not decisive for an interpretation in 
 the context of absorption.
This is because Seyfert 2 AGNs with similar absorbing 
 column densities show higher as well as lower equivalent widths 
 (Guainazzi et al. \cite{Guainazzi2005}). 

The spectra can be described with a power law continuum absorbed by 
 two different absorbers, one ionised and the other only 
 partially covering the source. 
Here two geometrical scenarios are possible: either the two absorbers 
 are located behind each other or the two absorbers only 
 partially cover the continuum source.  
A description of the spectra in the two states requires a change of 
 the covering fraction of the neutral material as well as a change 
 of the ionisation parameter. 
Therefore, 
 absorbing material has to move with respect to the 
 continuum source and our line of sight.

Although the discussed absorption scenarios provide statistically 
 acceptable descriptions of the spectra there are several arguments 
 against this interpretation:
\begin{itemize}
\item The 3-6 keV excess emission above the continuum power-law observed 
 during the deep minimum state is not well described as there are always 
 systematic residua left, - compare Figure~\ref{Fig1}. 
\item The residua at energies $>$6 keV in the rest frame of the source 
 can be reduced if the continuum emission is described with a power 
 law with $\Gamma=3.5$.  
 However, such a steep continuum has no parallel in other 
 AGNs (Piconcelli et al. \cite{Piconcelli2005})
 and as such is very unlikely to explain the observation.
\item Another possibility would be to explain the behaviour at energies 
 $>$6 keV with absorption similar to that discussed by Boller et al. 
 (\cite{Boller2002}) for 1H 0707-49.  
 A more general recent discussion of the topic can be found in
  Boller (\cite{Boller2006}). 
 However, with this assumption we introduce a further, third, absorption component.
\item But most important is that in all scenarios the ionisation parameter of 
 the deep minimum state is estimated to be significantly higher than the ionisation 
 parameter found for the high state, in sharp contrast to the 
 expectation that a higher source flux should cause a higher ionisation state.
\end{itemize}
All together the arguments are against an interpretation of the spectra 
 uniquely in terms of absorption.

In view of these we tested models that are composed of a power law continuum 
 and an X-ray ionised reflection on the accretion disk.
These models lead to statistically excellent descriptions of the data and 
 especially allow us to describe both states in a consistent way. 
There are two issues requiring further discussion: the ionisation parameter 
 of the warm absorber and the disappearance of the power law continuum 
 component during the deep minimum state.

An interpretation in the sense of reflection on an accretion disk leads 
 to values for the ionisation parameters of the absorber which are much more 
 consistent with values commonly found for AGNs.
But most important is the fact that the change of the ionisation 
 parameter is consistent with the variation of the flux. 
Comparing the continuum flux normalization measured for the high state of 
 the source with the 90\% upper limit of the continuum normalization obtained 
 for the deep minimum state (fit 24) we expect a drop of the ionisation parameter 
 by a factor of 100. 
From the measured ionisation parameters we obtain a ratio of 27$^{+420}_{-23}$. 
The change of the ionisation parameters therefore goes qualitatively in the 
 expected direction. 
Even quantitatively the measured ratio of the ionisation parameter is in 
 agreement with the value expected from the change of the 
 continuum flux. 
For comparison none of the absorption scenarios was able to provide only 
 qualitatively a consistent picture. 
As the larger reflection fraction during the deep minimum may imply a higher 
 ionization of the reflector we modelled both states simultaneously with the 
 ionization of the reflector as free parameter (fit 25). 
Unfortunately, the event statistics obtained are insufficient to measure any 
 real difference. 
Finally we verified that the best estimate for the ionisation parameter of the 
 deep minimum is a global minimum by stepping the parameter through the 
 allowed range. 

The joint modelling of both  spectral states of the source spectra leads to a description 
 of the deep minimum state in which the power law continuum component 
 completely vanishes, i.e. the whole observed spectrum is interpreted 
 as emission from X-ray ionised reflection on an accretion disk.
Although it is  unexpected to have AGN spectra without the 
paradigmatic power law component, we note that Miniutti and Fabian 
(\cite{Miniutti2004}) predicted exactly this situation.

In order to explain observed uncorrelated variability between direct 
 continuum and reflection components including the iron line, Miniutti 
 and Fabian (\cite{Miniutti2004}) explored a light-bending model. 
They assumed a primary source of X-rays located close to a central, 
 maximal rotating Kerr black hole and illuminating both the observer 
 at infinity and the accretion disk. 
The authors showed that even in the case of intrinsic constant luminosity 
 the observed flux can vary by more that an order of magnitude as the 
 height of the primary source above the accretion disks varies due to 
 strong light bending. 
Miniutti and Fabian identified three different regimes 
 (named regime I, II and III) according to the correlation between the 
reflection-dominated component and the direct continuum. 

The deep minimum state of \object{PG~2112+059} can be interpreted as an observation in the 
 regime I in which the observed direct continuum is very low.  
In this regime the primary source is at low height such that the primary 
 source emission suffers from strong light bending that dramatically reduces 
 the power law continuum at infinity (Miniutti \& Fabian \cite{Miniutti2004}).  
In addition Miniutti and Fabian  predicted for regime I 
 a very broad iron line emission profile without any prominent ``horn''-feature 
 (compare Figure 4 in Miniutti \& Fabian \cite{Miniutti2004}). 
The observed 3-6 keV ``bump'' during the deep minimum state is qualitatively in agreement 
 with the prediction.

There are several other AGNs for which absorption versus relativistically 
 blurred X-ray ionised reflection on an accretion disk is discussed. 
Examples are 
  \object{1H 0707-495} 
 (Boller et al. \cite{Boller2002}, 
  Fabian et al. \cite{Fabian2002}, 
  Tanaka et al. \cite{Tanaka2004}, 	
  Fabian et al. \cite{Fabian2004}) or 
 \object{1H 0419-577}
 (Fabian et al. \cite{Fabian2005},
  Pounds et al.  \cite{Pounds2004a},
  Pounds et al.  \cite{Pounds2004b}).
But in all these cases an absorption scenario is much more plausible than 
 in the case of \object{PG~2112+059}. 
A deep minimum state with the complete disappearance of the continuum power 
 law was reported for only three other AGNs: 
  \object{NGC 4051} (Uttley et al. \cite{Uttley1999}),  
  \object{NGC 2992} (Gilli et al. \cite{Gilli2000}) and 
  \object{NGC 1365} (Risaliti et al. \cite{Risaliti2007}). 
In all these cases the authors report narrow iron line emission with equivalent 
 widths of $>$0.7 keV, i.e. values which are above the 3-$\sigma$ 
 upper-limits obtained here for \object{PG~2112+059}.
Consequently in all cases the authors explain the deep minimum state with 
 reflection on material distant to the supermassive black hole 
 (e.g. broad-line region in the case of  \object{NGC 1365}). 
The appearance of the deep minimum is either explained by absorbing 
 material moving into the line of sight (\object{NGC 1365}) or a break down of the 
 accretion process and primary X-ray emission 
 (\object{NGC 2992} and  \object{NGC 4051}). 
Therefore, \object{PG~2112+059} appears to be the best 
 example of an AGN where the effect of light-bending 
 close to the central black hole has been detected so far.

A better statistical basis is required to finally decide on the nature of
  the X-ray emission of \object{PG~2112+059}. 
To achieve this objective we have successfully applied for a 220 ks long observation 
of \object{PG~2112+059} with XMM-Newton which is planned to be only performed if 
the quasar is found in the deep minimum state. 
If we can obtain such a spectrum and if the current interpretation 
 (a relativistially blurred ionised X-ray reflection emitted from an 
 accretion disk with the power law continuum suppressed, or even not present, due to light 
 blending) holds, then there are several quite remarkable consequences:
\begin{itemize}
 \item 
The spectra and the associated variability pattern will allow us to test 
 the predictions of the light-bending model (Miniutti \& Fabian \cite{Miniutti2004}). 
Such data will allow us to compare the emission by 
 an accretion disk near the supermassive black hole with theoretical predictions 
with unprecedented accuracy. 
\item
The question arises weather the weakness of other X-ray weak quasars is caused 
 by light bending, too. 
A hint might be provided by  \object{PG~1535+547}, which shows a very similar spectral 
 and temporal behaviour (Ballo et al. \cite{Ballo2007}).  
The modelling of several XMM-Newton spectra of X-ray weak quasars lead to warm 
 absorbers with very high ionisation parameters 
 (e.g. Schartel et al. \cite{Schartel2005}, for \object{PG~1535+547} and \object{PG 1001+054}; 
 Brinkmann et al. \cite{Brinkmann2004}, for \object{PG~1411+442} and 
 Piconcelli et al. \cite{Piconcelli2004a}, for \object{Mrk 304},
  Boller et al. \cite{Boller2002}, Fabian et al. \cite{Fabian2002}, 	
 Tanaka et al. \cite{Tanaka2004}, and Fabian et al. \cite{Fabian2004} for
 \object{1H 0707-495} ).  
Similar to \object{PG~2112+059}, in these cases the high ionisation parameters might 
 be a hint of a more complex spectral scenario and especially of X-ray 
 ionised reflection on accretion disks. 
The statement that absorption is the primary cause of the 
 weakness of (soft) X-ray weak quasars 
 (Brandt, Laor and Wills \cite{Brandt2000}) might be
  challenged. 
In addition the question about the nature of quasar types that are 
 closely related to X-ray weak quasars, like BAL quasars   
 (Brandt, Laor and Wills \cite{Brandt2000}),  would be raised, too.
\item
Many observations of quasars in the X-ray region aim to find features that 
 originate near to the central supermassive black hole. 
Due to the still very limited effective area of the current generation of X-ray 
 telescopes a strong selection for X-ray bright sources usually is required 
 to keep the exposure time reasonably long. 
 But this might be a bias, as the selection favours a high flux of 
 the continuum power law with respect to the relativistic blurred X-ray ionised 
 reflection from the accretion disk, which has a very small equivalent 
 width and is hard to detect.
\item
Further spectra in combination with their temporal characteristics might even 
 open the possibility of testing the general theory of relativity in the strong field 
 region with the current generation of X-ray satellites, as suggest 
 at the workshop ``Rethinking gravity'' (Hogan \cite{Hogan2007}).
\end{itemize}

\section{Conclusions}
\label{Con}

\begin{enumerate}
 \item
 In November 2005 \object{PG~2112+059} was in a deep 
  minimum state where the flux was lower by a factor 
  of 10 in comparison to the May 2003 data.
 \item 
 An interpretation of the spectral changes in the context 
 of an absorption scenario, although statistically 
 acceptable, is unsatisfying: the modelling with a neutral and 
 an ionised absorber still leaves systematic residua, 
 results in high column densities and very high ionisation 
 parameters. 
 Especially, the ionisation parameters found for the deep 
 minimum are higher than the ionisation parameters found for the 
 high state, implying either the presence of physically different absorbers 
 or a significant movement of the absorbers relative to 
 the X-ray sources.
 \item 
  An interpretation in the context of relativistically blurred 
 X-ray ionised reflection on an accretion disk near the 
 supermassive black hole behind a warm absorber provides a 
 plausible scenario for the spectral changes. 
 In this case the deep minimum state shows no continuum component 
 which can be explained by light bending 
 (Miniutti \& Fabian \cite{Miniutti2004}). 
 In this scenario we obtain ionisation parameter values that are 
 typical of AGNs, and especially find for the deep minimum state 
 lower ionisation parameters than for the high state, as 
 expected for a static absorber.
 \item 
 Should further observations confirm the light-bending scenario 
 for \object{PG~2112+059} then the source will offer unique possibilities 
 to specifically test the light-bending model and the emission of 
 an accretion disk near the supermassive back hole in general.
 \item 
 The finding for \object{PG~2112+059} challenges the suggestion that 
 absorption is the primary cause of the weakness of 
 X-ray weak quasars. 
 \end{enumerate}

\begin{acknowledgements}
Based on observations obtained with XMM-Newton, an ESA science
 mission with  instruments and contributions directly funded by
 ESA Member States and NASA.

We thank the referee, Dr. Giovanni Miniutti, for the many useful comments. 

\end{acknowledgements}

\listofobjects

\end{document}